\documentclass[doublecol]{epl2} 

\usepackage{isotope}
\usepackage{caption}
\usepackage{subcaption}
\usepackage{url}
\usepackage{enumitem}
\usepackage{comment}
\usepackage{lineno}

\title{Experimental study of the 2n--transfer reaction \isotope[138]{Ba}(\isotope[18]{O},\isotope[16]{O})\isotope[140]{Ba}
in the projectile energy range 61--67~MeV}
\shorttitle{2n--transfer study of \isotope[140][]{Ba}} 

\author{
A.~Khaliel\inst{1}\thanks{email: achalil@phys.uoa.gr}
\and T.J.~Mertzimekis\inst{1}
\and F.C.L.~Crespi\inst{2}
\and G.~Zagoraios\inst{1}
\and D.~Papaioannou\inst{1}
\and N.~Florea\inst{3}
\and A.~Turturica\inst{3}
\and L.~Stan\inst{3}
\and N.~M\u{a}rginean\inst{3}
}
\shortauthor{A. Khaliel \etal}
\institute{
  \inst{1}
    Department of Physics, National \& Kapodistrian University of Athens, Zografou Campus, GR--15784, Athens, Greece\\
  \inst{2}
    Universit\'a degli Studi di Milano and INFN sez. Milano, Milano, Italy\\
  \inst{3}
    National Institute for Physics and Nuclear Engineering, Magurele, Romania
}
\pacs{25.45.Hi}{Transfer reactions}
\pacs{21.10.Tg}{Lifetimes,Widths}
\pacs{25.60.Pj}{Fusion reactions}

\abstract{
Two--neutron transfer reactions serve as an important tool for nuclear--structure
studies in the neutron--rich part of the nuclear chart. In this article, we report
on the first experimental attempt to populate the excited states of \isotope[140]{Ba}
employing the 2n--neutron transfer reaction
\isotope[138]{Ba}(\isotope[18]{O},
\isotope[16]{O})\isotope[140]{Ba}.
\isotope[140]{Ba} is highly important, as it is placed on the onset of octupole
correlations and the lifetimes of its excited states are completely unknown, with
the sole exception of the first $2^+$ state. The experiment was carried out at the
Horia Hulubei National Institute for Physics and Nuclear Engineering (IFIN--HH) in
Magurele, Romania. Lower limits on the lifetimes of ground state band up to the $8^+$
state are reported. Furthermore, relative cross sections regarding the 2n--transfer
reaction with respect to the fusion and the total inelastic reaction channels have
been deduced. Further investigation directions of the nuclear structure of
\isotope[140]{Ba} are also discussed.
}

\begin{document}

\maketitle

\section{Introduction}
\label{intro}
Multinucleon Transfer Reactions (MTR) are important tools for nuclear structure
studies~\cite{2007_Szilner,2008_Zagrebaev,Corradi_2009}. Especially for energies
close to the Coulomb barrier, transfer--reaction cross sections are a large fraction
of the total reaction cross section~\cite{Corradi_2009}, thus leading to a significant
population of excited states of the produced nuclei. It is under discussion that such
reactions can offer a new pathway for the study of neutron--rich transuranium isotopes
and superheavy elements~\cite{Heinz2018}, as their expected yields are comparable to
the fusion reactions, while providing the advantage of offering a wide range of
populated isotopes during the same experiment.

Two--neutron transfer reactions have been successfully used for populating the excited
states of nuclei, which are moderately rich in neutrons~\cite{2017_Leoni_PRL}. Recently,
the neutron-rich $^{144,146}$Ba isotopes were studied experimentally in terms of their
$B(E3)$ values~\cite{Bucher2016,Bucher2017} using radioactive beams and Coulomb
excitation. The respective $B(E3)$ values, although featuring large uncertainties,
were found to be significantly larger than any theoretical prediction. Thus, a study
of \isotope[140]{Ba} is important for establishing the onset of octupole correlation,
as well as assessing the degree of collectivity in the barium isotopic chain as a
function of neutron number. In addition, the lifetimes of the lower--lying states
of \isotope[140]{Ba} are unknown, with the sole exception of the first $2^+_1$ state,
as reported in~\cite{Bauer2012}.

Cross--section data, either absolute or relative, are important for estimating the
degree of level populations of the reaction products. Experimental cross section data
are still scarce for such reactions, especially for barium. Barium is a material
that oxidizes very quickly when exposed to air, thus making the manufacturing of
a target quite challenging and a ``difficult'' nucleus to study using stable beams.

In this work, we report on the relative cross sections of the 2n--transfer reaction
\isotope[18]{O}+\isotope[138]{Ba}$\rightarrow$\isotope[16]{O}+\isotope[140]{Ba}
with respect to the fusion evaporation reaction
\isotope[18]{O}+\isotope[138]{Ba}$\rightarrow$\isotope[152]{Gd}+4n,
as well as to the total inelastic channel. These ratios can serve as a reference
point for the theoretical studies, i.e. refining Optical Model Potentials, or
further experimental studies using such reactions. Furthermore, lower limits on
lifetimes of the observed ground--state band states are reported by taking
into consideration the limitations of the Doppler Shift Attenuation Method
(DSAM)~\cite{Alexander1978,Petkov_1999} for the particular system.
\begin{figure}[!ht]
    \centering
    \includegraphics[width=0.4\textwidth]{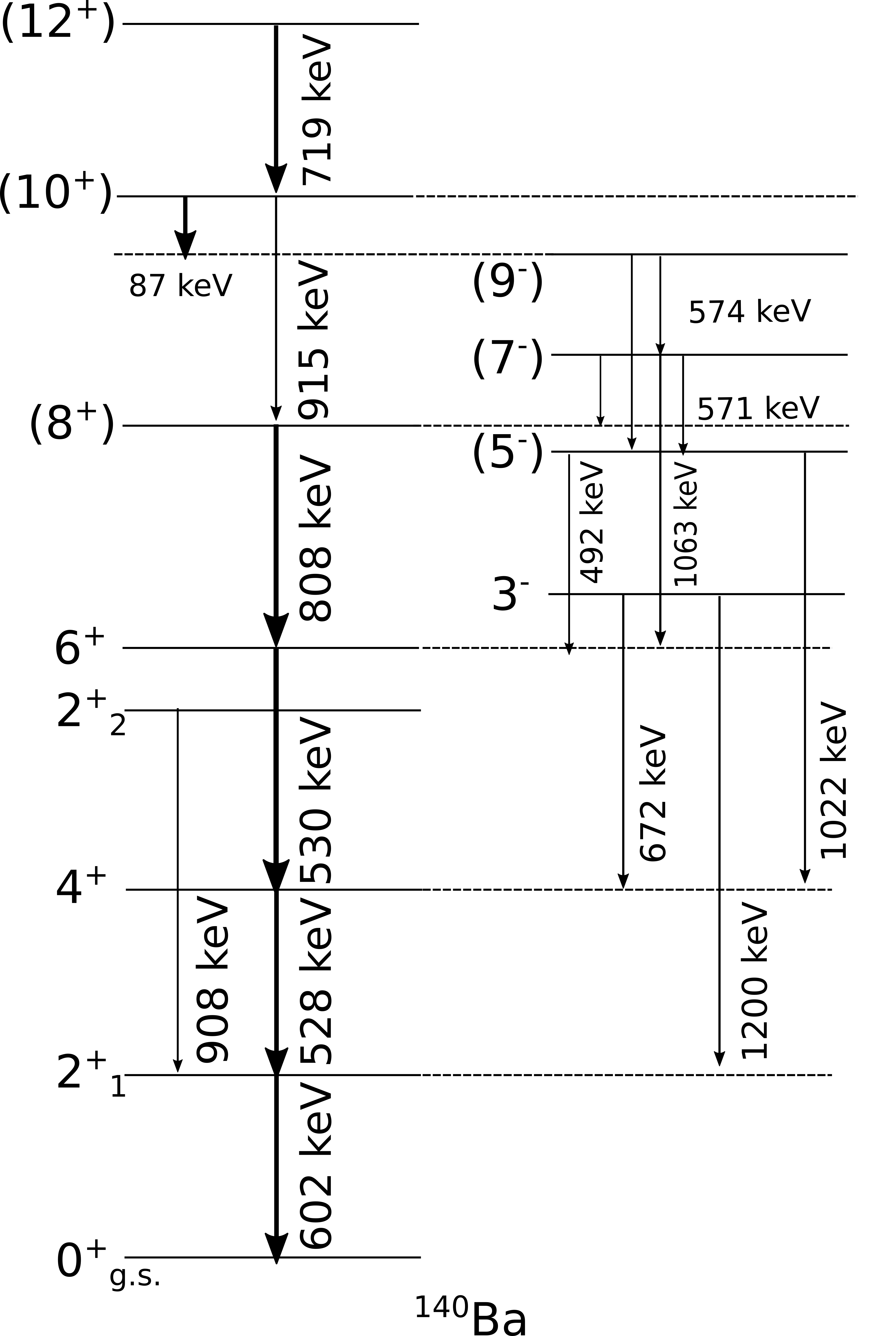}
    \caption{
    Partial level scheme of \isotope[140]{Ba}, showing the ground state band
    and a side band~\cite{NNDC}. The alternating parity of the states of two
    bands is a hint for significant octupole correlations.
    }
    \label{fig:level_scheme}
\end{figure}

\section{Experimental Details}

\subsection{Experimental setup}

The experiment was carried out at the 9~MV Tandem Accelerator Laboratory at the
Horia Holubei National Institute of Physics and Nuclear Engineering (IFIN--HH)
in Magurele, Romania. Four projectile energies were studied near the Coulomb
barrier of the reaction, namely 61, 63, 65 and 67~MeV. The subsequent $\gamma$ decay
was detected by the ROSPHERE array~\cite{BUCURESCU20161} loaded with 15 HPGe detectors.

\subsection{Target manufacturing}

\begin{figure*}[ht]
\centering
\includegraphics[width=\textwidth, keepaspectratio]{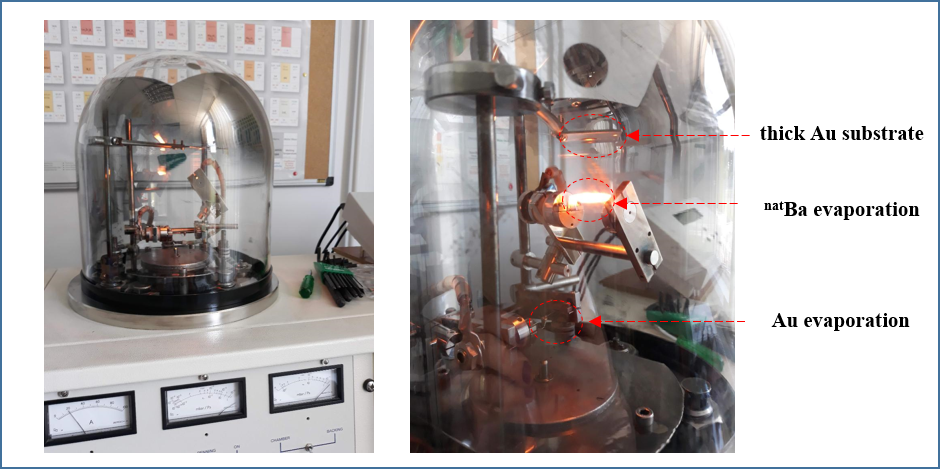}
\caption{
The evaporator chamber (left) and a picture taken during the evaporation procedure (right).
}
\label{fig:target}
\end{figure*}

The manufacturing of the \isotope[nat]{Ba} target in metallic form presents important
difficulties, as it is a material that oxidizes extremely fast. Therefore, as it is illustrated in Fig.~\ref{fig:target}, a gold--sandwiched \isotope[nat]{Ba} target was
prepared in the Target Preparation Laboratory of IFIN--HH~\cite{Florea2015}. The target
exhibits the following structure:
Au (4.88~mg\,cm$^{-2}$) / \isotope[nat]{Ba} (2~mg\,cm$^{-2}$) / Au (0.5~mg\,cm$^{-2}$).
The metallic \isotope[nat]{Ba} layer  (abundance of $^{138}$Ba = 71.7\%) was obtained
through the metalothermic reduction reaction of BaCO$_3$ with La metal powder as reducing
agent. For this purpose about double the stoichiometric amount of La metal powder was
thoroughly ground with the calculated amount of BaCO$_3$, in an agate mortar set. The
resulted mixture was pressed into a pellet, which was inserted into a pinhole tantalum
boat. Both ends of the boat were fixed to the high current electrodes of the Quorum
technologies E6700 bench top evaporator device. The gold foil of 4.88~mg\,cm$^{-2}$
thickness, prepared in advance by rolling, was glued to the target frame and placed
4~cm above the tantalum boat in the evaporator. After a high vacuum of
$3.5\times10^{-5}$~mbar was reached, a low current was applied through the tantalum
boat to degassing of CO$_2$ resulted from the thermal decomposition of BaCO$_3$.
Therefore, the current through the tantalum boat was slightly increased until the
reduction temperature was reached. The evaporation process was carried out until the
desired thickness was obtained. The obtained \isotope[nat]{Ba} layer was covered with
a thin gold layer of 0.5~mg\,cm$^{-2}$ without breaking the vacuum, to protect the
metallic \isotope[nat]{Ba} against oxidation. This deposition was made with a tungsten
basket, fixed at 9~cm distance above the substrate. 

The determination of the thick gold backing's thickness was done by weighing,
while the other two layers were determined by calculating the thickness from
the initial amount of the substance used.

\section{Analysis and Results}

\subsection{Lifetimes}

\begin{figure}[!ht]
    \centering
    \begin{subfigure}{0.45\textwidth}
    \centering
    \includegraphics[width=\textwidth]{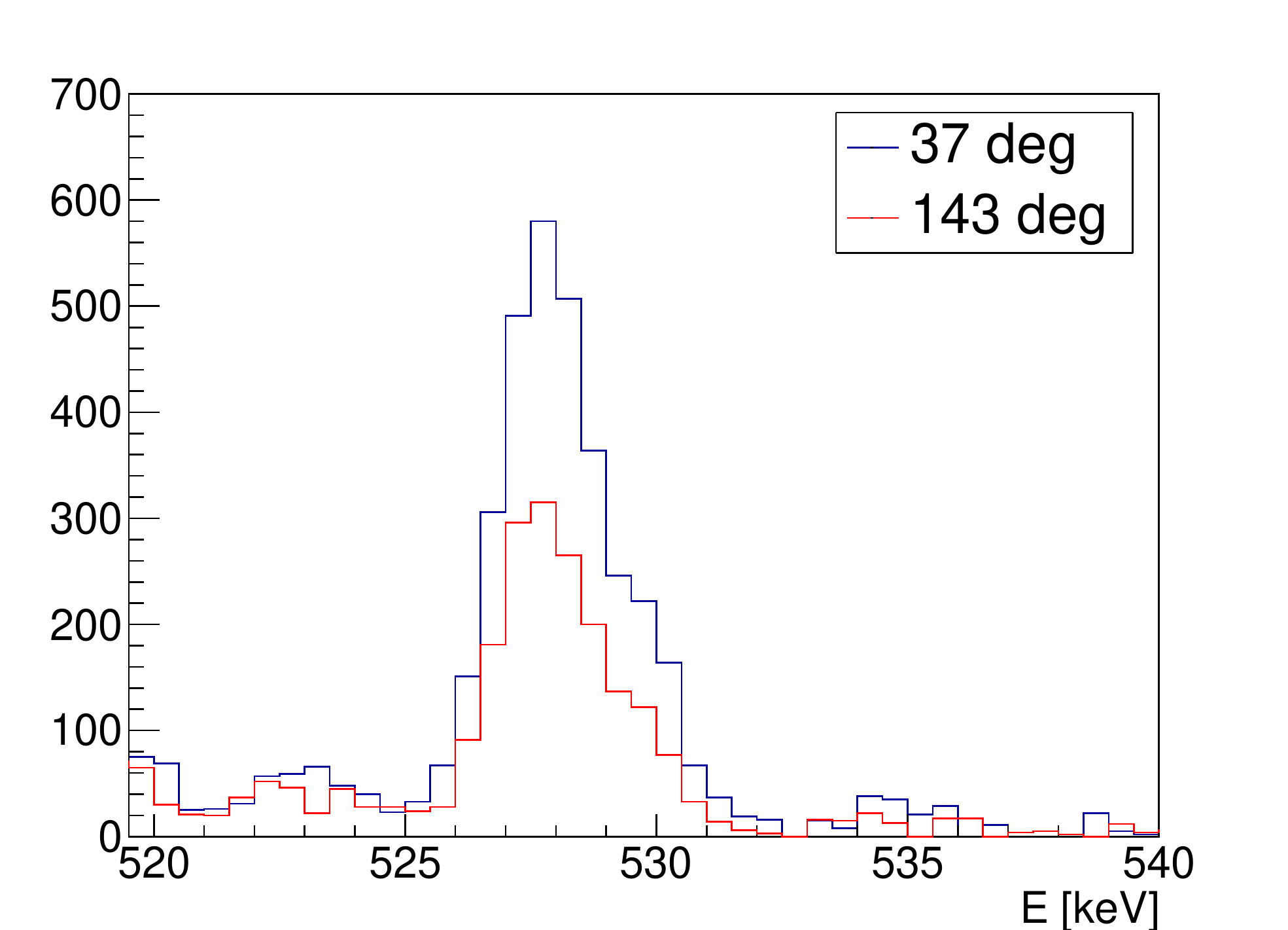}
    \caption{}
    \label{subfig:528 peak}
    \end{subfigure}%
    \\
    \begin{subfigure}{0.45\textwidth}
    \centering
    \includegraphics[width=\textwidth]{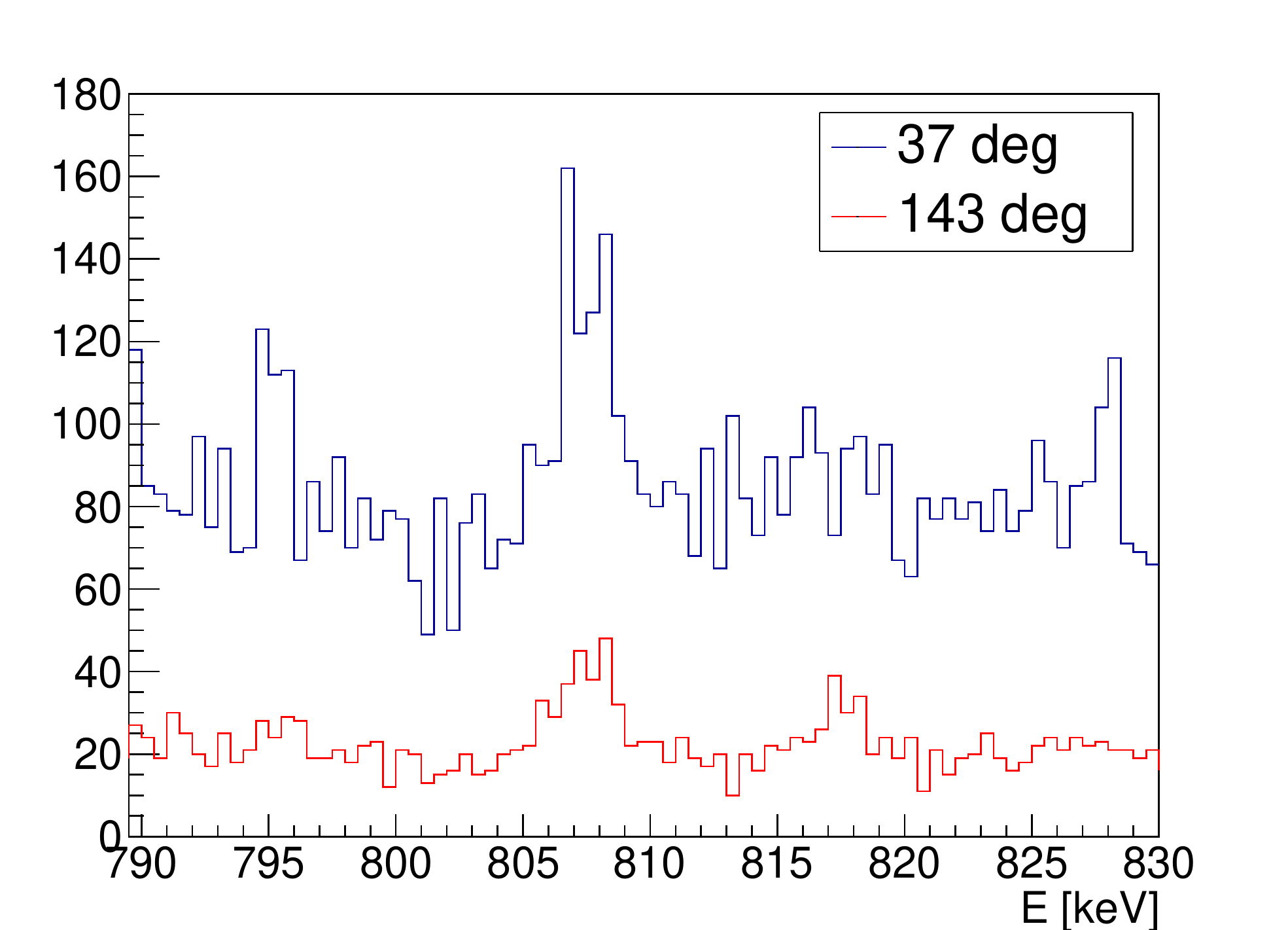}
    \caption{}
    \label{subfig:808 peak}
    \end{subfigure}
    \caption{Backward (143$^\circ$) and forward (37$^\circ$) spectra for
    (a) the 528 and 530~keV overlapping transitions, and (b) for the
    808~keV transition. The spectra show no backward--forward lineshapes.}
    \label{fig:backward forward}
\end{figure}

Lower limits on lifetimes of the states up to $8^+$ in the ground--state
band~\cite{NNDC}, corresponding to the observed transitions can be set,
by taking into account the limitation of the Doppler Shift Attenuation Method
(DSAM). In Fig.~\ref{subfig:528 peak}, the two overlapping transitions of
energies 528 and 530~keV are shown, depopulating the $4^+$ and $6^+$ states
of the ground--state band. As it can be seen for the spectra recorded in the
backward (143$^\circ$) and forward ring (37$^\circ$), no visible lineshapes
can be distinguished. The same holds for Fig.~\ref{subfig:808 peak}, where
the transition of 808~keV is depopulating the $8^+$, also in the ground--state
band.
 
The maximum recoil velocity in the particular reaction mechanism is 2\% the
speed of light. At such recoil velocities, the range of lifetimes that can be
measured with DSAM should be lower than approximately 1~ps \cite{Alexander1978, Petkov2006_564, Mihai2010, 2018_Petkov}. The present limit is established in terms
of the range of lifetimes that the particular method can be applied, and not the
sensitivity of the experimental setup.

\subsection{Relative cross sections}

The cross section of a reaction can be estimated by the relation:
\begin{equation}
    \sigma = \frac{N_R}{\Phi N_t}
\end{equation}
where $N_R$ is the number of occurring reactions, $N_t$ is the number of
target nuclei that the beam interacts with, and $\Phi$ is the incident flux
of projectiles.

The reactions
\begin{eqnarray*}
\isotope[18]{O}+\isotope[138]{Ba}&\rightarrow&\isotope[16]{O}+\isotope[140]{Ba}\\
~&\rightarrow&\isotope[152]{Gd}+4n\\
~&\rightarrow&\isotope[18]{O}+\isotope[138]{Ba}^*
\end{eqnarray*}
stem from the same entry channel and occur inside the barium foil. In general,
the relative cross section for two different exit channels $\alpha$, $\beta$
can be estimated by:
\begin{equation}
\sigma_R = \frac{N_R(\alpha)}{N_R(\beta)}
\label{e:sigma_rel}
\end{equation}
i.e. by determining the ratios of the corresponding number of reactions.

In the present case, the number of reactions can be deduced by measuring all observed
photopeaks feeding the ground state of the produced nuclei for the two above exit
channels, and then correcting with the full absolute efficiency of the ROSPHERE array:
\begin{equation}
    N_R = \frac{A}{\epsilon_{abs}}
\end{equation}
where A is the area of the photopeak and $\epsilon_{abs}$ is the absolute efficiency.

\begin{figure*}[!ht]
    \centering
    \includegraphics[width=0.95\textwidth]{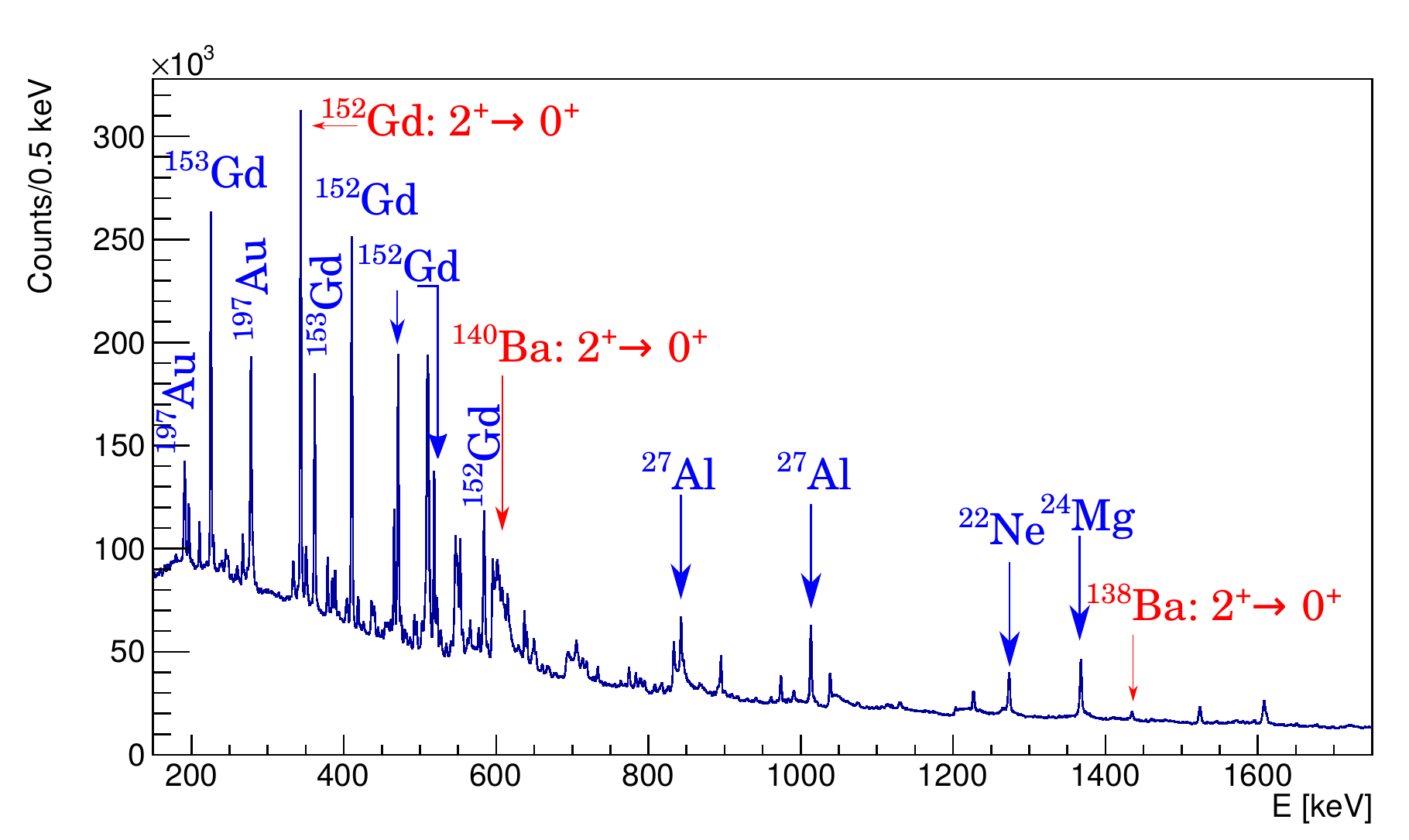}
    \caption{
    Total projection spectrum from the $\gamma-\gamma$ matrix acquired using
    the ROSPHERE array. Transitions in barium isotopes are marked, as well as
    several from the fusion--evaporation channels. A few contaminant peaks are
    also indicated.}
    \label{fig:65MeV_37deg}
\end{figure*}

For the 2n--reaction
\isotope[18]{O}+\isotope[138]{Ba}$\rightarrow$\isotope[16]{O}+\isotope[140]{Ba},
only the $E_\gamma = 602$~keV transition was observed, while for the
\isotope[18]{O}+\isotope[138]{Ba}$\rightarrow$\isotope[152]{Gd}+4$n$
reaction, only the transition with $E_\gamma = 344$~keV was recorded in the data
(see Fig.~\ref{fig:level_scheme}). Also, for the total inelastic channel, the
transition $E_\gamma=1436$~keV was observed. A projection spectrum of the full
$\gamma-\gamma$ matrix is shown in Fig.~\ref{fig:65MeV_37deg}, with the peaks of
interest marked.

\begin{figure}[!hb]
    \centering
    \begin{subfigure}{0.45\textwidth}
    \centering
    \includegraphics[width=\textwidth]{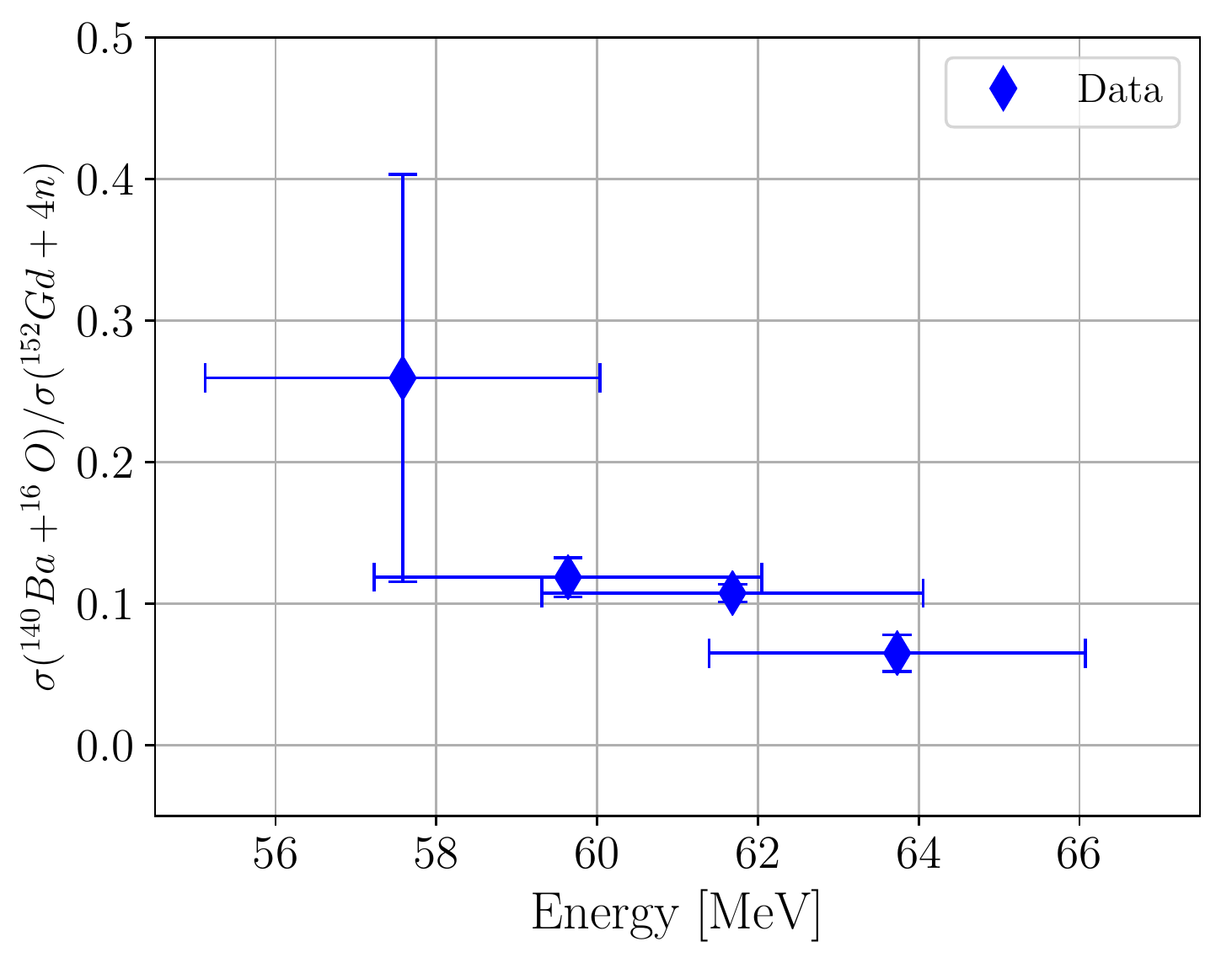}
    \caption{}
    \label{subfig:602_peak_cnt_ratio}
    \end{subfigure}%
    \\
    \begin{subfigure}{0.45\textwidth}
    \centering
    \includegraphics[width=\textwidth]{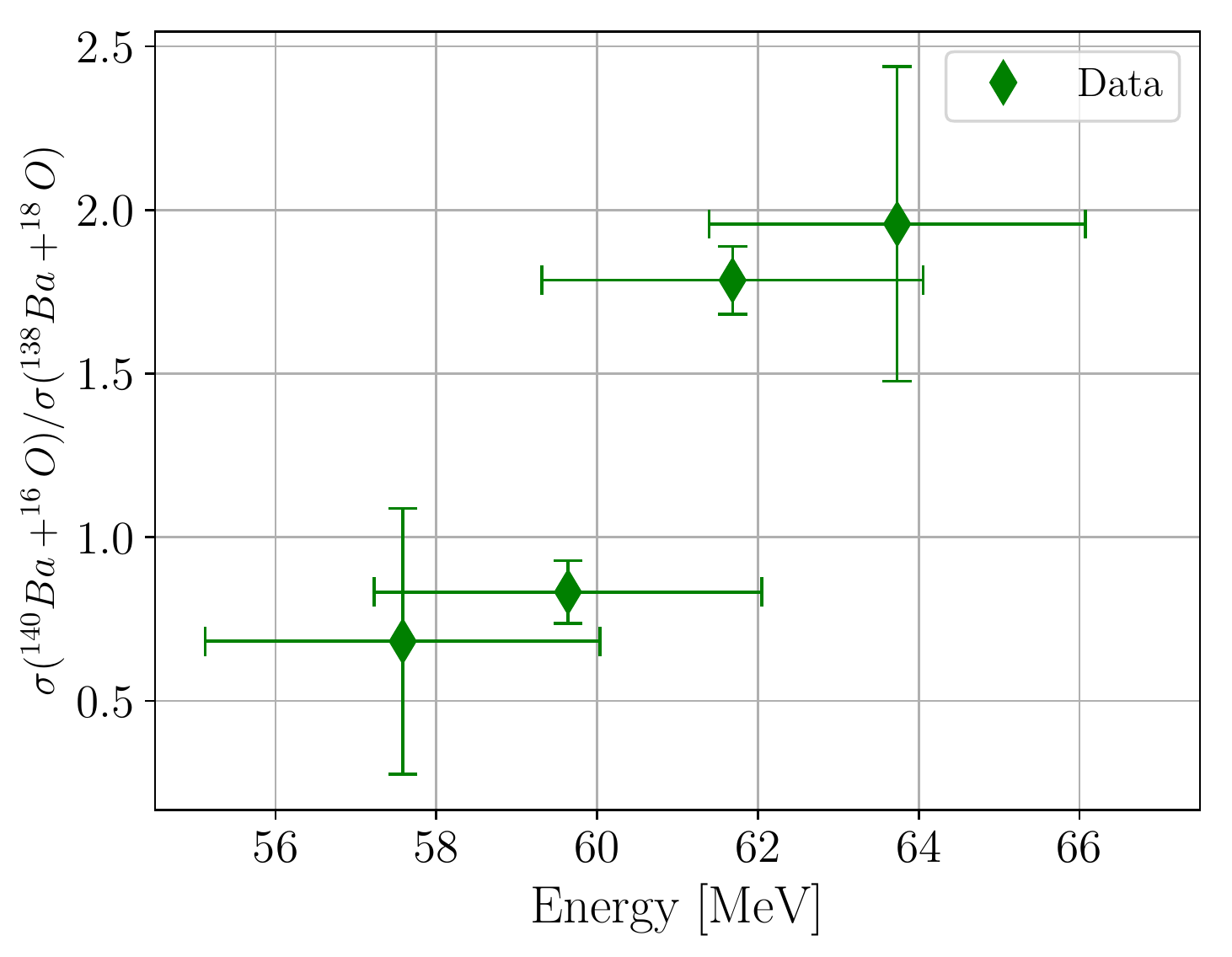}
    \caption{}
    \label{subfig:528_peak_cnt_ratio}
    \end{subfigure}
    \caption{Relative cross section of the two neutron--transfer reaction
    \isotope[18]{O}+\isotope[138]{Ba}$\rightarrow$\isotope[16]{O}+\isotope[140]{Ba}
    with respect to the fusion evaporation reaction
    \isotope[18]{O}+\isotope[138]{Ba}$\rightarrow$\isotope[152]{Gd}+4$n$ (left)
    and the ratio with respect to the total inelastic channel
    \isotope[18]{O}+\isotope[138]{Ba}$\rightarrow$\isotope[16]{O}+\isotope[138]{Ba}$^*$ (right).}
    \label{fig:rel_cross}
\end{figure}

By extracting the ratios, and taking into account the energy loss inside the barium
foil of the target (Table~\ref{tab:summary}), the results for the relative cross
sections of the 2n--neutron transfer reaction
\isotope[18]{O} + \isotope[138]{Ba}$\rightarrow$\isotope[16]{O} + \isotope[140]{Ba}
with respect to the fusion--evaporation counterpart
\isotope[18]{O} + \isotope[138]{Ba}$\rightarrow$\isotope[152]{Gd} + 4n and with
respect to the total inelastic channel are shown in Fig.~\ref{fig:rel_cross} as a
function of energy in the laboratory system.

\begin{table*}[!ht]
    \centering
    \caption{
    Experimental results and theoretical calculations. From left to right column:
    $E_b$ is the beam energy; $E_{Ba}$ is the incident energy of at the front of the Barium foil, by taking into consideration the beam energy loss inside the front Au foil; 
    $\Delta E_{Ba}$ is the total energy loss in the barium foil;
    $E_{eff}$ is the effective energy at the middle of the barium foil;
    $\sigma_R^{fus}$ is the ratio of the 2n--transfer reaction cross section over
    the cross section of the fusion-evaporation channel (see eq.~\ref{e:sigma_rel});
    $\sigma_R^{inel}$ is the ratio of the 2n--transfer reaction cross section over
    the total inelastic cross section;
    $\sigma_{Gd}$ and $\sigma_{Ba}$ are normalized cross sections (see text for details); 
    $\sigma_{PACE}$ and $\sigma_{GRAZING}$ are results of \texttt{PACE4} and \texttt{GRAZING~9} calculations for the \isotope[18]{O} + \isotope[138]{Ba}$\rightarrow$\isotope[152]{Gd} + 4n and \isotope[18]{O} + \isotope[138]{Ba}$\rightarrow$\isotope[16]{O} + \isotope[140]{Ba} reactions, respectively.
    All values are in the laboratory system.}
    \resizebox{\textwidth}{!}{  
    \begin{tabular}{cccccccccc}
    \hline
    $E_{b}$ & $E_{Ba}$ & $\Delta E_{Ba}$ & $E_{eff}$ & $\sigma_R^{fus}$ & $\sigma_R^{inel}$ & $\sigma_{Gd}$ &
    $\sigma_{Ba}$ & $\sigma_{PACE4}$ & $\sigma_{GRAZING}$ \\
    (MeV) & (MeV) & (MeV) & (MeV) & & & (mb) & (mb) & (mb) & (mb)\\ \hline \hline
    61 & 60.04 &4.91 & 57.59 & 0.3$\pm$0.1 & 0.7$\pm$0.4 & 0.43$\pm$0.01 & 0.11$\pm$0.06 & -- & 0.1112 \\
    63 & 62.05 & 4.83 & 59.64 & 0.12$\pm$0.01 & 0.8$\pm$0.1 & 2.46$\pm$0.01 & 0.29$\pm$ 0.04 & 0.00178 & 0.33087 \\
    65 & 64.06 & 4.75 & 61.69 & 0.107$\pm$0.006  & 1.8$\pm$0.1 & 8.02$\pm$0.04 & 0.86$\pm$0.06 & 0.847 & 0.63257 \\
    67 & 66.07 &4.68 & 63.73 & 0.06$\pm$0.01 & 2.0$\pm$0.5 & 23.4$\pm$0.1 & 1.5$\pm$0.3 & 23.4 & 0.89368 \\
    \hline
    \end{tabular}
    }
    \label{tab:summary}
\end{table*}

\section{Cross section predictions}

In order to further investigate the experimental values of the relative cross
sections measured in the present work, theoretical calculations have been performed
using the
\texttt{GRAZING~9}~\cite{Winther1994} and
\texttt{PACE4}~\cite{Gavron_1980_PhysRevC.21.230} codes.

The former uses Winther's grazing model~\cite{Winther1994}, which has been proven
successful for the description of one-- or two--nucleon transfer reactions~\cite{Reisdorf1994}.
Calculations performed with the \texttt{GRAZING~9} code use a semi--classical
approach developed in Ref.~\cite{Wendt1988}. For a small number of nucleon transfers
(up to 6--8 neutrons)  and for nuclei close to the magic shell closures, the particular
model describes experimental data very well; however, it tends to slightly underestimate
the data for large numbers of nucleons (see discussion in~\cite{Samarin2013}).

On the other hand, the \texttt{PACE4} code is the latest version of a modified JULIAN
code~\cite{1980_JULIAN} and uses the Bass model~\cite{Bass1980_PhysRevLett.39.265},
which was derived by using a geometric interpretation of available experimental data
combined with a Monte--Carlo approach to determine the decay of the compound system
in the framework of Hauser--Feshbach formalism~\cite{Hauser1952_PhysRev.87.366}.
As stated in Ref.~\cite{Reisdorf1994}, the Bass model potential provides an
overall excellent description for the fusion cross sections at energies starting
from the Coulomb barrier and above. However, experimental evidence shows that the
particular model significantly underestimates the cross section data below the
Coulomb barrier~\cite{Reisdorf1994}. 

\begin{figure}[!ht]
    \centering
    \begin{subfigure}{0.45\textwidth}
    \centering
    \includegraphics[width=\textwidth]{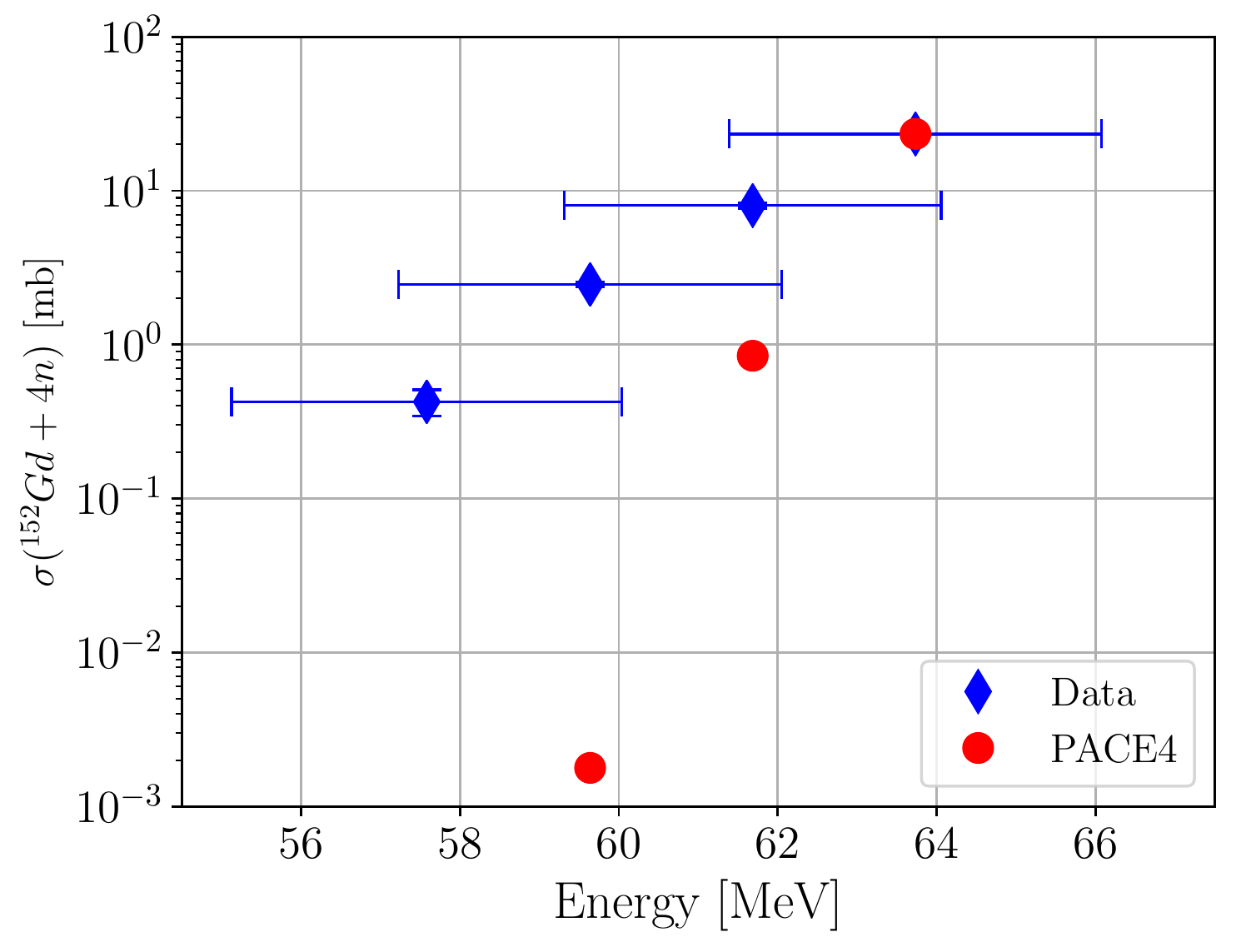}
    \caption{}
    \label{subfig: 152Gd xsection}
    \end{subfigure}%
    \\
    \begin{subfigure}{0.45\textwidth}
    \centering
    \includegraphics[width=\textwidth]{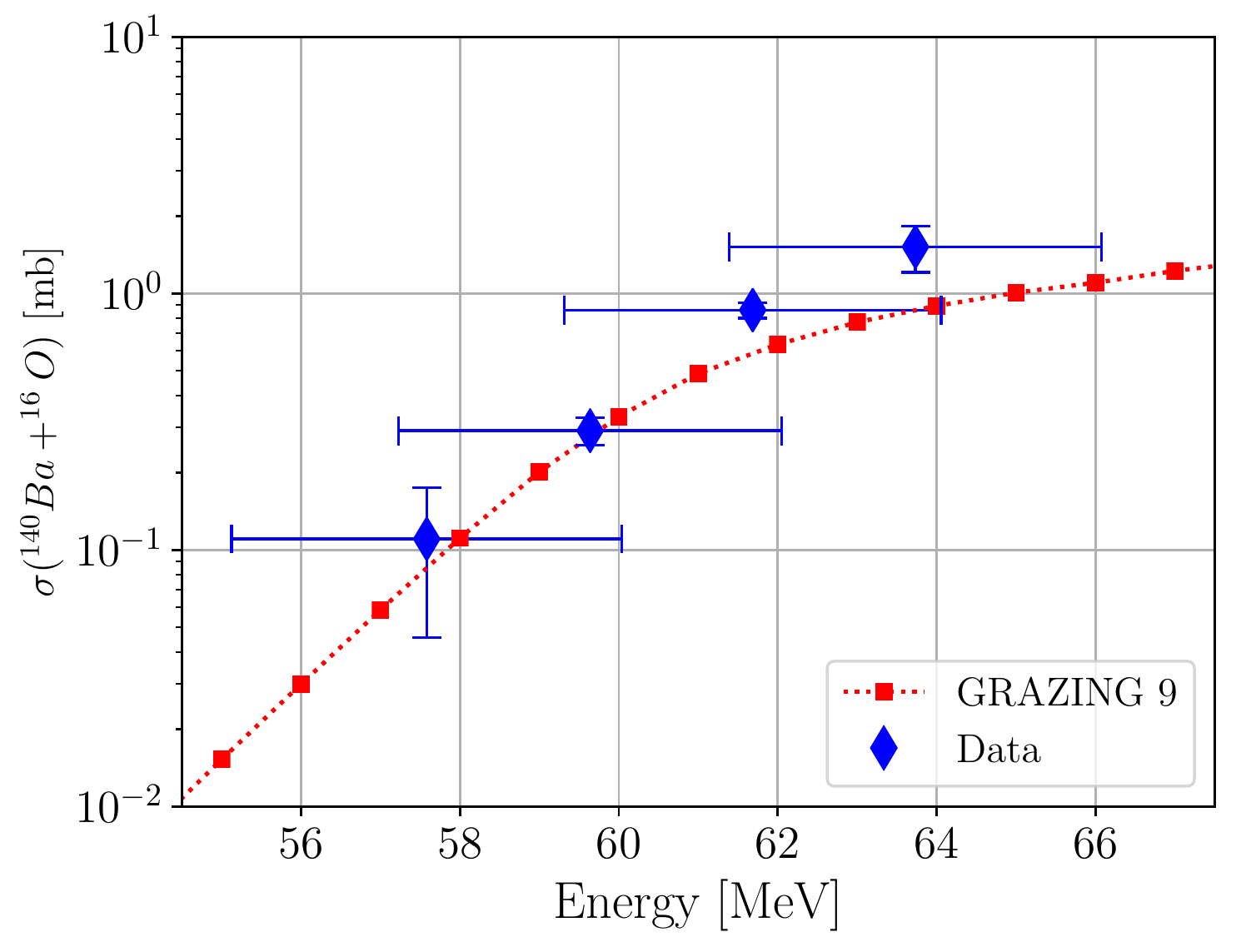}
    \caption{}
    \label{subfig:Ba xsection}
    \end{subfigure}
    \caption{
    (a) Normalized cross sections of the fusion--evaporation channel after
    normalization to \texttt{PACE4} calculations (see text). Vertical error bars
    are smaller than the symbol size.
    (b) Deduced cross sections for the 2n--neutron transfer channel (solid diamonds) 
    after taking into account the results from panel (a), together with
    \texttt{GRAZING~9} calculations (solid squares). No scaling is involved in any
    of the data sets in panel (b). The dotted line is to guide the eye. Scales
    of y--axes in (a) and (b) are different.
    }
    \label{fig:cross_sections}
\end{figure}

All calculations have been performed using the default parameters each code employs.
Fig.~\ref{fig:cross_sections} shows the results of the calculations: Fig.~\ref{fig:cross_sections}a includes cross section calculations with
\texttt{PACE4} (solid circles), while Fig.~\ref{fig:cross_sections}b contains
calculated cross sections with \texttt{GRAZING~9} (solid squares). \texttt{PACE4}
could not produce a value at the lowest energy ($E_{eff}$=57.6~MeV), as was expected,
since this value is far below the barrier. The experimental absolute yield for the
fusion--evaporation channel is shown in Fig.~\ref{fig:cross_sections} (solid diamonds).
All data points have been normalized with a single numerical factor. That factor
was estimated by scaling the experimental point at the energy nearest to the
barrier ($E_{eff}$=63.7~MeV) with the respective value calculated with \texttt{PACE4}
(see overlapping points in Fig.~\ref{fig:cross_sections}a).

Calculations with \texttt{GRAZING~9} are shown in Fig.~\ref{fig:cross_sections}b
(solid squares). In the same graph, experimental data of the 2n--transfer reaction
\isotope[18]{O} + \isotope[138]{Ba} $\rightarrow$ \isotope[16]{O} + \isotope[140]{Ba}
are shown. These data have been extracted from the ratio in Fig.~\ref{fig:rel_cross}
taking into account the scaled cross sections of the fusion--evaporation channel,
as described earlier. No further scaling is involved.

\section{Discussion and future investigations}

Within the present framework, a study of the nucleus \isotope[140]{Ba} by using
the $2n$--transfer reaction
\isotope[18]{O} + \isotope[138]{Ba} $\rightarrow$ \isotope[16]{O} + \isotope[140]{Ba},
has been performed. By considering the kinematics of the reaction studied and the
limitation of DSAM, lower limits on the lifetimes of 3 states of the ground state
band have been set over 1 ps. Of course, further studies are necessary in order to
further constrain the above limit. The present results also sets the path for using
a different technique for the measurement of the particular lifetimes, such as the
plunger technique or the fast--timing technique. For direct measurement of the reduced
transition probabilities, especially for the $B(E3)$ corresponding to the first
$3^-$ state, the use of radioactive beams and Coulomb excitation technique can override
a lot of issues, such as possible target contamination and the level population strength.

The relative cross sections between the 2n--transfer reaction
\isotope[18]{O} + \isotope[138]{Ba} $\rightarrow$ \isotope[16]{O} + \isotope[140]{Ba}
and the competing fusion--evaporation reaction
\isotope[18]{O} + \isotope[138]{Ba} $\rightarrow$ \isotope[152]{Gd} + 4n
have been deduced by taking into account the relative yield of the two observed
transitions feeding the ground state of the two produced nuclei. The relative cross
section behavior seems to follow a reducing pattern with respect to beam energy,
showing that the fusion--evaporation channel becomes stronger faster, as the Coulomb
barrier is approached. This behavior is rather expected given the fact that the
reactions occur in the pure--tunneling energy range.
In addition, the relative cross sections of the reaction
\isotope[18]{O} + \isotope[138]{Ba} $\rightarrow$ \isotope[16]{O} + \isotope[140]{Ba}
and the total inelastic channel are presented. The inelastic channel is a competing
reaction channel, which as can be seen from Fig.~\ref{fig:rel_cross}b, the respective
cross section values are of the same order of magnitude within the studied energy
range. The behavior of these cross sections follows an increasing pattern, indicating
that the 2n--transfer reaction shows a stronger increase as the energy increases
towards the Coulomb barrier. 

Absolute cross sections deduced with this method, taking advantage of the ratios
of cross sections, may be lower than their actual values, as some transitions feeding
the ground state may not be observed in the spectra,resulting in missing strength
in the overall estimation. In addition, the $^{152}$Gd decay features $E0$ transitions
directly to the ground state. It is very hard to observe such transitions in the
$\gamma$--spectrum, despite their contribution to the total number of produced nuclei.
While this can be usually treated as a weak effect, it cannot be assumed with certainty.

Calculations with the theoretical codes \texttt{PACE4} and \texttt{GRAZING~9} have
been performed to provide a comparison with experimental data produced in the present
work. The scaling of the experimental data to the \texttt{PACE4} result at the maximum
energy, almost identical to the energy of the barrier, can be trusted to produce
absolute cross sections for the absolute cross sections in the fusion--evaporation
channels. This becomes evident when the deduced cross sections for the studied
2n--transfer reaction are further compared to \texttt{GRAZING~9} calculations. There
is a very good agreement between experimental data and theory, both in trend and
in magnitude. The two lowest energy points are effectively the same within the
experimental uncertainty, while the discrepancy between the rest is of the order of
20\%. It has to be stressed again that this comparison involves no other scaling
than the one used for the fusion--evaporation channel. It would be interesting also
to check this method in more detail for energies near and above the barrier in the
future, especially for neighboring nuclei in this mass regime.

In conclusion, the results provide useful information for the specific case study,
for both the experimental cross sections, as well as the validity of theoretical
models at energies near the barrier. 2n--transfer reactions are a very useful tool
to study unstable, moderately neutron--rich nuclei. To this end, the knowledge of
the 2n--transfer--to--fusion cross section ratio can be extremely useful, for example,
for reducing the fusion background, especially in nuclear structure studies. In
addition, cross section ratios can help in constraining the optical model potential
phenomenological parameters, for the better understanding of systems involving
heavy--ion reactions. Such experimental data in the region around \isotope[140]{Ba}
are very scarce, but also very important for studies trying to extend our knowledge
on more exotic species towards the neutron--dripline.

\acknowledgements
This research work was supported by the Hellenic Foundation for Research and Innovation (HFRI) and the General Secretariat for Research and Technology (GSRT) under the HFRI PhD Fellowship Grant (GA. No. 74117/2017). Partial support from ENSAR2 (EU/H2020 project number: 654002) is acknowledged.
\bibliographystyle{eplbib}
\bibliography{barium_database}

\end{document}